\documentclass[10pt,letterpaper,superscriptaddress,pra,preprint]{revtex4-1}
\usepackage[latin1]{inputenc}
\usepackage{hyperref}

\usepackage{graphicx,color}

\newcommand{\qtext}[1]{{\color{magenta}#1}}
\renewcommand{\qtext}[1]{{\color{magenta}}}

\begin{document}

\title{Certified quantum non-demolition measurement of material systems}

%
%


\newcommand{\ICFOAddress}{ICFO -- Institut de Ciencies Fotoniques, Av. Carl Friedrich Gauss, 3, 08860 Castelldefels, Barcelona, Spain}
\newcommand{\ICREAAddress}{ICREA -- Instituci\'{o} Catalana de Re{c}erca i Estudis Avan\c{c}ats, 08015 Barcelona, Spain}
\newcommand{\CavendishAddress}{Cavendish Laboratory, University of Cambridge, JJ Thomson Avenue, Cambridge CB3 0HE, UK}
\newcommand{\OptosAddress}{Optos, Carnegie Campus,
Dunfermline, KY11 8GR, Scotland, UK}

\author{Morgan~W.~Mitchell}\address{\ICFOAddress}\address{\ICREAAddress}
\author{Marco~Koschorreck}\address{\CavendishAddress}
\author{Marcin~Kubasik}
\address{\OptosAddress}
\author{Mario~Napolitano}\address{\ICFOAddress}
\author{Robert~J.~Sewell}\address{\ICFOAddress}


\email{morgan.mitchell@icfo.es} \homepage{http://mitchellgroup.icfo.es}



\begin{abstract}
An extensive debate on quantum non-demolition (QND) measurement, reviewed in Grangier et al. [Nature, {\bf 396}, 537 (1998)], finds that true QND  
measurements must have both non-classical state-preparation capability and non-classical information-damage tradeoff.  Existing figures of merit for these  {non-classicality} criteria require direct measurement of the signal variable and are thus difficult to apply to optically-probed material systems.  Here we describe a method to demonstrate both  criteria without need for to direct signal measurements.  Using a covariance matrix formalism and a general noise model, we compute meter observables for QND measurement triples, which suffice to compute all QND figures of merit.   The result will allow certified QND measurement of atomic spin ensembles using existing techniques.
\end{abstract}



\maketitle

\newcommand{\be}{\begin{equation}}
\newcommand{\ee}{\end{equation}}
\newcommand{\bea}{\begin{eqnarray}}
\newcommand{\eea}{\end{eqnarray}}
\newcommand{\sinc}{\mbox{\rm sinc}}
\newcommand{\ba}{{\bf a}}
\newcommand{\bb}{{\bf b}}
\newcommand{\bd}{{\bf d}}
\newcommand{\bk}{{\bf k}}
\newcommand{\bq}{{\bf q}}
\newcommand{\bJ}{{\bf J}}
\newcommand{\bj}{{\bf j}}
\newcommand{\bE}{{\bf E}}
\newcommand{\bx}{{\bf x}}
\newcommand{\br}{{\bf r}}
\newcommand{\bH}{{\bf H}}
\newcommand{\bB}{{\bf B}}
\newcommand{\bP}{{\bf P}}
\newcommand{\bC}{{\bf C}}
\newcommand{\bD}{{\bf D}}
\newcommand{\bv}{{\bf v}}
\newcommand{\brho}{{\bf \rho}}
\newcommand{\balpha}{\stackrel{\leftrightarrow}{\alpha}}
\newcommand{\twovector}[2]{\left[\begin{array}{c}#1\\#2\end{array}\right]}
\newcommand{\twomatrix}[4]{\left[\begin{array}{cc}#1&#2\\#3&#4\end{array}\right]}
\newcommand{\threevector}[3]{\left[\begin{array}{c}#1\\#2\\#3\end{array}\right]}
\newcommand{\threematrix}[9]{
\left[\begin{array}{ccc}#1&#2&#3\\#4&#5&#6\\#7&#8&#9\end{array}\right]}
\newcommand{\braket}[2]{\left<#1|#2\right>}
\newcommand{\ket}[1]{\left|#1\right>}
\newcommand{\bra}[1]{\left<#1\right|}
\newcommand{\Schrodinger}{Schr\"{o}dinger~}
\newcommand{\FigWidth}{15 cm}
\newcommand{\expect}[1]{\left<#1\right>}
\newcommand{\adag}{{a^\dagger}}
\newcommand{\Tr}{{\rm Tr}}
\newcommand{\chione}{\chi^{(1)}}
\newcommand{\chitwo}{\chi^{(2)}}
\newcommand{\chithr}{\chi^{(3)}}
\newcommand{\bee}{{\bf e}}
\newcommand{\bA}{{\bf A}}
\newcommand{\bnab}{{\bf \nabla}}
\newcommand{\epzero}{\varepsilon_{0}}
\newcommand{\exercise}{\hspace{-.25in}\Large}
\newcommand{\hH}{\hat H}
\newcommand{\hn}{\hat n}
\newcommand{\ha}{\hat a}
\newcommand{\hd}{\hat d}
\newcommand{\hx}{\hat x}
\newcommand{\hp}{\hat p}
\newcommand{\hadagger}{\hat a^{\dagger}}
\newcommand{\hb}{\hat b}
\newcommand{\hbdag}{\hat b^{\dagger}}
\newcommand{\hau}{\hat{\underline{a}}}
\newcommand{\haudagger}{\hau^{\dagger}}
\newcommand{\hbA}{\hat b_{A}}
\newcommand{\hbE}{\hat b_{E}}
\newcommand{\hE}{\hat {\bf E}}
\newcommand{\hnbE}{\hat {E}}
\newcommand{\hnbD}{\hat {D}}
\newcommand{\hEp}{\hat {E}^{(+)}}
\newcommand{\hEm}{\hat {E}^{(-)}}
\newcommand{\hX}{\hat {X}}
\newcommand{\tbE}{{\bf\cal E}}
\newcommand{\tbF}{{\bf\cal F}}
\newcommand{\Ep}{E^{(+)}}
\newcommand{\Em}{E^{(-)}}
\newcommand{\tbX}{{\bf\cal X}}
\newcommand{\tP}{{\cal P}}
\newcommand{\tbP}{{\bf\cal P}}
\newcommand{\tE}{{\cal E}}
\newcommand{\tF}{{\cal F}}
\newcommand{\tG}{{\cal G}}
\newcommand{\tX}{{\cal X}}
\newcommand{\cE}{{\cal E}}
\newcommand{\cF}{{\cal F}}
\newcommand{\cD}{{\cal D}}
\newcommand{\cP}{{\cal P}}
\newcommand{\nn}{\nonumber \\ }
\newcommand{\nnn}{\nonumber \\ &  &}
\newcommand{\nne}{\nonumber \\ & = &}
\newcommand{\nnequiv}{\nonumber \\ & \equiv &}
\newcommand{\nnapprox}{\nonumber \\ & \approx &}
\newcommand{\nnpropto}{\nonumber \\ & \propto &}
\newcommand{\nnp}{\nonumber \\ & & +}
\newcommand{\nnm}{\nonumber \\ & & -}
\newcommand{\nnt}{\nonumber \\ & & \times}
\newcommand{\bdp}{\bd^{(+)}}
\newcommand{\bdm}{\bd^{(-)}}
\newcommand{\bEp}{\bE^{(+)}}
\newcommand{\bEm}{\bE^{(-)}}
\newcommand{\bBp}{\bB^{(+)}}
\newcommand{\bBm}{\bB^{(-)}}
\newcommand{\hain}{\ha_{\rm IN}}
\newcommand{\haout}{\ha_{\rm OUT}}
\newcommand{\hadaggerin}{\hadagger_{\rm IN}}
\newcommand{\hadaggerout}{\hadagger_{\rm OUT}}
\newcommand{\hEin}{\hnbE_{\rm IN}}
\newcommand{\hEout}{\hnbE_{\rm OUT}}
\newcommand{\haone}{\ha_1}
\newcommand{\hadaggerone}{\hadagger_1}
\newcommand{\hatwo}{\ha_2}
\newcommand{\hadaggertwo}{\hadagger_2}
\newcommand{\hainone}{\ha_{{\rm IN},1}}
\newcommand{\hadaggerinone}{\hadagger_{{\rm IN},1}}
\newcommand{\haintwo}{\ha_{{\rm IN},2}}
\newcommand{\hadaggerintwo}{\hadagger_{{\rm IN},2}}
\newcommand{\curl}{\nabla \times}
\newcommand{\ve}{\varepsilon}
\newcommand{\Adagger}{A^{\dagger}}
\newcommand{\Bdagger}{B^{\dagger}}
\newcommand{\Cdagger}{C^{\dagger}}
\newcommand{\Ddagger}{D^{\dagger}}
\newcommand{\Lag}{{\cal L}}
\newcommand{\var}{{\rm var}}

\newcommand{\arctanh}{\rm{arctanh}}

\newcommand{\Pe}{P_e}
\newcommand{\Pg}{P_g}
\newcommand{\PeZ}{P_{e0}}
\newcommand{\PgZ}{P_{g0}}

%
%
%

\newcommand{\Green}{{\cal G}}
\newcommand{\GreenBack}{{\cal H}}
\newcommand{\Eff}{{Q}}
\newcommand{\CorrFun}{{H}}
\newcommand{\BVConst}{{\beta}}
\newcommand{\Source}{{\cal S}}

\newcommand{\sQ}{p}
\newcommand{\sV}{q}
\newcommand{\sT}{r}
\newcommand{\SQ}{P}
\newcommand{\SV}{Q}
\newcommand{\ST}{R}
\newcommand{\SP}{D}  
\newcommand{\Proj}{\Pi}  
\newcommand{\bQ}{{\bf \SQ}}
\newcommand{\bV}{{\bf \SV}}
\newcommand{\bT}{{\bf \ST}}

\newcommand{\SQx}{\SQ_{x}}
\newcommand{\SQy}{\SQ_{y}}
\newcommand{\SQz}{\SQ_{z}}
\newcommand{\SVx}{\SV_{x}}
\newcommand{\SVy}{\SV_{y}}
\newcommand{\SVz}{\SV_{z}}
\newcommand{\STx}{\ST_{x}}
\newcommand{\STy}{\ST_{y}}
\newcommand{\STz}{\ST_{z}}
\newcommand{\sQx}{\sQ_{x}}
\newcommand{\sQy}{\sQ_{y}}
\newcommand{\sQz}{\sQ_{z}}
\newcommand{\sVx}{\sV_{x}}
\newcommand{\sVy}{\sV_{y}}
\newcommand{\sVz}{\sV_{z}}

\newcommand{\SQV}{C}
\newcommand{\SJQV}{T}
\newcommand{\bSV}{{\bf\SV}}
\newcommand{\bSQ}{{\bf\SQ}}
\newcommand{\bST}{{\bf\ST}}
\newcommand{\bSQV}{{\bf\SQV}}
\newcommand{\bSJQV}{{\bf\SJQV}}

\newcommand{\cov}{{\rm cov}}
\newcommand{\ScIn}{{\bar{Q}_x}}
\newcommand{\C}{\tilde{C}}
\newcommand{\CondVar}{E[\var(J_z)|\SQ_y]}

\newcommand{\LTrans}{r_L}
\newcommand{\AStay}{r_A}

\newcommand{\nnd}{\nonumber \\ & & \div}

\newcommand{\bS}{{\bf S}}

\section{Introduction}
A quantum non-demolition (QND) measurement is one which
provides information about a quantum variable while leaving
it unchanged and accessible for future measurements.
The approach was originally suggested as a means to avoid
 measurement back-action in gravitational wave
  detection \cite{braginsky75,ThornePRL1978,UnruhPRD1979,
  CavesRMP1980,BraginskyS1980}.  QND measurements
of optical fields both provided the first demonstration{s} and 
led to a considerable refinement of the
  understanding of QND measurements in practice  \cite{GrangierN1998}.
 More recently, QND measurements have been employed to prepare spin-squeezed atomic
  states 
  \cite{kuzmich99,Takano2009,Appel2009,Schleier-Smith2010a,
  Chen2011} and with nano-mechanical systems \cite{RuskovPR2005}.

In a generic QND measurement, a `meter' and a `system' variable interact via a selected
Hamiltonian.  The meter can then be directly measured to gain indirect information about the system.   In the context of optical QND measurements, the question of when a measurement should be considered QND has been much discussed (see  \cite{GrangierN1998} for references).   Two distinct non-classicality criteria emerge:  A state preparation criterion requires small uncertainty in the system variable after the measurement while a second criterion describes the information-damage tradeoff in the measurement.  While some operations such as filtering or optimal cloning can be non-classical in one or the other criterion, a true QND measurement is non-classical in both \cite{GrangierN1998}.  

With the aid of figures of merit \cite{holland90,RochAPBLO1992,grangier92} describing the quantum-classical boundary, optical QND measurements satisfying both criteria have been demonstrated 
\cite{BencheikhPRL1995, BencheikhPRL1997, BruckmeierPRL1997, GoobarPRL1993, LevensonPRL1993, PereiraPRL1994, PoizatPRL1993, RochAPBLO1992, RochPRL1993, RochPRL1997}.  These figures of merit make use of the fact that the optical signal beam, after the QND measurement, can be verified by a direct, i.e., destructive, measurement with quantum-noise-limited sensitivity.  Typically such a direct measurement is not available in atomic QND.  Rather, repeated QND measurement has been used to show the state preparation criterion \cite{Takano2009, Appel2009, Schleier-Smith2010a, Chen2011, SewellARX2011} by conditional variance measurements.   Here we show how repeated QND measurements can also be used to test the information-damage tradeoff, and thus to certify full QND performance without direct access to the system variable.

\section{Model}

As in the pioneering work by Kuzmich, {\em et al.}
\cite{kuzmich98,kuzmich99}, we consider the collective spin of an
atomic ensemble, described by the vector angular momentum operator
$\bJ$.   We note that a variety of other physical situations are described in the same 
way, e.g. by using a pseudo-spin to describe a clock transition \cite{Appel2009}.  
The optical polarization of any probe pulse is described by a vector Stokes operator $\bS$
 \be S_i \equiv \frac{1}{2} \ba^\dagger \sigma_i \ba, \ee $i=x,y,z$ where
$\sigma_i$ are the Pauli
 matrices, $\ba \equiv \{a_+,a_-\}^T$ and $a_\pm$ are
 annihilation operators for circular-plus and circular-minus polarizations.   

We define Stokes operators $\bf \SQ,\SV$ for the first and second
pulses, respectively. The operators $\bJ,\bSQ,\bSV$ each obey the
angular momentum commutation relation $ \left[ L_x, L_y \right] =
i  L_z$ and cyclic permutations (for simplicity, we take $\hbar = 1$).   For notational convenience,
we define the combined optical variables $\bSQV \equiv \bSQ \oplus
\bSV$ and the total variable $\bSJQV \equiv \bJ \oplus \bSQV$. We
will be interested in the average values of these operators, which
we write as $\bar{\bJ} \equiv \expect{\bJ}$ and similar, and the
covariance matrices, which we write as \be \label{Eq:CovMat}
\tilde{J} \equiv \frac{1}{2} \expect{\bJ \wedge \bJ +(\bJ \wedge
\bJ)^T} - \expect{\bJ} \wedge \expect{\bJ} \ee and similar.   Our
approach follows that of Madsen and M{\o}lmer \cite{Madsen:2004,KoschorreckJPBAMOP2009}.

\newcommand{\supin}{^{(\rm in)}}
\newcommand{\supout}{^{(\rm out)}}

We assume that the input probe pulses are polarized as
$\bar{\SQ}_x\supin = \bar{\SV}_x\supin = \bar{S}_x\supin$ and
that the other average components are zero. We take the initial
covariance matrix for the system to be \be \tilde{T}_0 = \tilde{J}
\oplus \C
\ee  This form of the covariance matrix allows for
arbitrary prior correlations (including correlated technical
noise) among the two optical pulses, but no prior correlations
between the atoms and either optical pulse.

The interaction is described by an effective Hamiltonian \be \label{Eq:Hamiltonian}
H_{\rm eff} = g J_z S_z, \ee where $g$ is a constant
\cite{echaniz08}.  
%
{This} QND interaction, to lowest order in $g \tau$, where $\tau$ is
the interaction time of the pulse and atoms, produces a rotation
of the state, $\bSJQV\supout =\bSJQV\supin -i
\tau[\bSJQV\supin,H_{\rm eff}]$. This has the effect of
imprinting information about $J_z$ on the light without changing
$J_z$ itself: \bea \label{Eq:QNDIntEqs}
S_y\supout &=& S_y\supin + \kappa' S_x\supin J_z\supin \\
J_z\supout &=& J_z\supin.\eea Here $\kappa' = g \tau$ and $\bS$
is $\bSQ$ or $\bSV$ depending on which pulse-atom interaction is
being described.  The rotation can be described by a linear
transformation $ \bSJQV\supout = M_\SQ \bSJQV\supin $ (and thus
$ \tilde{\SJQV}\supout = M_\SQ \tilde{\SJQV}\supin M_\SQ^T $)
where $M_\SQ$ is equal to the identity matrix, apart from the
elements $(M_P)_{2,6} = \kappa' \bar{J}_x\supin$, and
$(M_P)_{5,3} = \kappa' \bar{S}_x\supin$. For later convenience,
we define $\kappa \equiv \kappa' \bar{S}_x\supin = g \tau
\bar{S}_x\supin $.

The effect of the second pulse is described by the matrix $M_\SV =
X_{} M_\SQ X_{} $ where 
\be X_{}\equiv \left(
\begin{array}{ccc}
1 & 0 & 0   \\
0 & 0 & 1   \\
0 & 1 & 0
\end{array}
\right) \otimes I_3
\ee

exchanges the roles of $\SQ$ and $\SV$, and $I_3$ is the
$3\times 3$ identity matrix.


\section{Reduction of uncertainty by QND measurement}

We first consider the case in which the interaction does not introduce
additional noise (although both the input atomic and optical states may be noisy).  
After interaction with the
first pulse, but before the arrival of the second pulse, the state
is described by $ \tilde{\SJQV}_{\SQ} \equiv M_\SQ \tilde{\SJQV}_0
M_\SQ^T $.
A component $\SQy$ of the first pulse is measured. Formally, this
corresponds to projection along the axis ${\bf m}_\SQ \equiv
\{0,0,0,0,1,0,0,0,0 \}^T$, and $\tilde{\SJQV}_{\SQ}$ is reduced to
\be \tilde{\SJQV}_{\SQ\SP} = \tilde{\SJQV}_{\SQ} -
\tilde{\SJQV}_{\SQ} (\Proj_Q \tilde{\SJQV}_{\SQ} \Proj_Q)^{\rm MP}
\tilde{\SJQV}_{\SQ}^T = \tilde{\SJQV}_{\SQ} -
\tilde{\SJQV}_{\SQ} \Proj_Q \tilde{\SJQV}_{\SQ}^T / {\rm
Tr}[\Proj_Q \tilde{\SJQV}_{\SQ}] \ee where $\Proj_Q \equiv {\bf
m}_\SQ \wedge {\bf m}_\SQ $ is the projector describing the
measurement and $()^{\rm MP}$ indicates the Moore-Penrose
pseudo-inverse.

We can directly calculate the resulting variance of $J_z$,  \be
\label{Eq:VarJzIdeal} \CondVar \equiv
(\tilde{\SJQV}_{\SQ\SP})_{3,3} =
\tilde{J}_{3,3}\frac{\C_{2,2}}{\kappa^2 \tilde{J}_{3,3} +
\C_{2,2}} .\ee This has a natural interpretation: The variance of
the detected projection $\SQ_y$ has two contributions:
$\kappa^2\tilde{J}_{3,3}$ from the atomic signal and $\C_{2,2}$
from the pre-existing optical noise.  $\tilde{J}_{3,3}$ is reduced
by the factor $1/(1+{\rm SNR})$ where SNR is the signal-to-noise
ratio of the measurement.  A similar result is found in reference
\cite{Madsen:2004}.  This post-measurement variance of the signal
variable describes the state-preparation capability of the QND 
measurement.  
{Absent} the ability to directly measure $J_z$, we must look
for observables which contain this same information.  

\section{Observable correlations}

After interaction with both the first and second pulses, we have $
\tilde{\SJQV}_{\SQ\SV} \equiv M_\SV\tilde{\SJQV}_{\SQ} M_\SV^T $.
  This matrix contains the variances and correlations that are
  directly measurable, namely those of the two light pulses.  These are
  \bea
\var(\SQy) &=& \C_{2,2}+ \kappa^2 \tilde{J}_{3,3}  \\ \var(\SVy)
&=& \C_{5,5}+ \kappa^2\tilde{J}_{3,3}   \\ \cov(\SQy,\SVy) &=&
\C_{2,5}+ \kappa^2 \tilde{J}_{3,3} . \eea   We note that for
$\kappa=0$, e.g. if the atoms are removed, the values are \bea
\label{Eq:VarsNoAtoms}  \var_{\rm NA}(\SQy) &=& \C_{2,2} \\
\var_{\rm NA}(\SVy) &=&\C_{5,5} \\ \cov_{\rm NA}(\SQy,\SVy) &=&
\C_{2,5}.  \eea


We see that the state preparation capability can be expressed in
terms of measurable quantities as \be \CondVar = \tilde{J}_{3,3}
\frac{\var_{\rm NA}(\SQ_y)}{\var(\SQ_y)},\ee which uses the
variance of the two measurements to determine the SNR.  Another
formulation, \bea \CondVar &=& \tilde{J}_{3,3} \frac{\var_{\rm NA}(\SQ_y)}
{ {\var_{\rm NA}(\SQ_y) + \cov(\SQy,\SVy)} - {\cov_{\rm
NA}(\SQy,\SVy)} }, \qtext{\rm~minus~inserted} \eea   expresses the residual variance
in terms of the atomic contribution to the correlation between
first and second pulses.

These simple expressions are only valid for noise-free interactions,
however.   In a real experiment, other effects are present which
introduce both noise and losses in the atomic and optical
variables. We now account for these other effects.


\section{General noise and loss}

We now consider noise produced in the atom-light interaction
itself, as well as losses.  The noise model we employ is very
general.  The
interaction of the first pulse with the atoms is described by \be
\label{Eq:GeneralQNDTransform}
 \tilde{\SJQV}_\SQ = M_\SQ \tilde{\SJQV}_0 M_\SQ^T + N_\SQ
 \ee
We assume that the coherent part of the interaction is $M_\SQ
\equiv \AStay I_3 \oplus \LTrans I_3 \oplus I_3 $ apart from
the elements $(M_P)_{2,6} = \kappa \bar{J}_x\supin$, and
$(M_P)_{5,3} = - \kappa \bar{S}_x\supin$.
%
%
%
Here $\AStay,\LTrans$ describe the fraction of atoms and photons,
respectively, that remain after the interaction. Thus $M_\SQ$
includes both the effect of $H_{\rm eff}$ and linear losses. We
leave $N_\SQ$ completely general, except that it does not affect
$\bSV$: $N_\SQ \equiv N \oplus 0 I_3$, where $N$ is a six-by-six
symmetric matrix.

Similarly, we describe interaction with the second pulse as
\be \label{Eq:GeneralVerifyTransform}
 \tilde{\SJQV}_{\SQ\SV} = M_\SV \tilde{\SJQV}_\SQ M_\SV^T + N_\SV
 \ee
 where $M_\SV = X M_\SQ X$ and $N_\SV = X N_\SQ X.$

Note that we assume that both the interaction $M$ and the noise
$N$ are the same for the first and second pulses (but act on
different variables, naturally).  This implies that optical
characteristics of the pulses such as detuning from resonance are
the same, a condition that can be achieved in experiments.  It
also assumes that the noise generated by the interaction is incoherent
and state-independent, as opposed to a more general, state-dependent noise
$N(\bJ,\bS)$. Nevertheless, in many situations $\bJ$ and $\bS$ are
nearly constant (only small quantum components change
appreciably), so that any reasonable $N(\bJ,\bS)$ would be
effectively constant.

As above, we can directly calculate $\tilde{\SJQV}_{\SQ\SP}$ and
$\tilde{\SJQV}_{\SQ\SV}$ to find \bea \CondVar &=&
\tilde{J}_{3,3}\AStay^2 +
{N_{3,3}}{}  -\frac{(\kappa r_A \tilde{J}_{3,3}+N_{3,5})^2 }
 { \kappa^2\tilde{J}_{3,3} +
\LTrans^2\C_{2,2} + N_{5,5} }  \eea and
\bea \var(\SQy) &=& \LTrans^2\C_{2,2}+\kappa^2\tilde{J}_{3,3} + N_{5,5} \\
\var(\SVy) &=& \LTrans^2\C_{5,5}+ \kappa^2
(\AStay^2\tilde{J}_{3,3} + N_{3,3})+  N_{5,5} \\ \cov(\SQy,\SVy)
&=& \LTrans^2\C_{2,5} + \kappa^2 (\AStay \tilde{J}_{3,3}+
N_{3,5}/ \kappa ). \eea Equation (\ref{Eq:VarsNoAtoms}) still
holds for the case with no atoms. We define
\bea \delta\var(\SQy) &\equiv& \var(\SQy) - \var_{\rm
NA}(\SQy)\LTrans^2 \\ \delta\var(\SVy) &\equiv& \var(\SVy) -
\var_{\rm NA}(\SVy)\LTrans^2 \hspace{8mm} \\ \delta\cov(\SQy,\SVy) &\equiv&
\cov(\SQy,\SVy)- \cov_{\rm NA}(\SQy,\SVy)\LTrans^2, \eea where the $r_L$ factors are included to account for atom-induced optical losses. 
It is then simple to check that 
 \bea \label{Eq:EffectivenessWLoss}
\CondVar&=&\tilde{J}_{3,3} + \kappa^{-2}\left(
\rule{0cm}{0.5cm}\delta\var(\SVy) - 
  \delta \var(\SQy) -
\frac{\delta\cov^2(\SVy,\SQy)}{\var(\SQy)} \right). \qtext{\rm~minus~inserted} \eea

We note that the QND measurement reduces the variance of $J_z$ if
the quantity in parentheses is negative, i.e., if \be
\label{Eq:SpinSqCondWLoss} \delta\cov^2(\SVy,\SQy)
> \var(\SQy) [\delta\var(\SVy)-\delta \var(\SQy) ]. \ee  Again,
there is an intuitive explanation: $\delta\cov(\SVy,\SQy)$, which
arises from the fact that both pulses measure the same atomic
variable $J_z$, is a measure of the atom-light coupling.
$[\delta\var(\SVy)-\delta \var(\SQy) ]$ expresses the difference
in atom-induced noise between the first and second pulses.  This
difference indicates a change in the atomic state, namely an
increase in $\var(J_z)$. The condition of Equation
(\ref{Eq:SpinSqCondWLoss}) compares these two effects and can be
tested knowing the statistics of the various measurements on $S_y$
and the optical transmission $\LTrans$.  The factors $\kappa^2,
\tilde{J}_{3,3}$ in equation (\ref{Eq:EffectivenessWLoss}) must be
determined by independent means.  For example, $\kappa$ can be
found by measuring the rotation of a state with known
$\expect{J_z}\ne 0$ and $\tilde{J}_{3,3}$ from the number of
atoms,  or the observed noise scaling of a 
known state \cite{KubasikPRA2009,KoschorreckPRL2010a}.

\section{Three-pulse experiments}

The above description of two-pulse experiments can be extended
straightforwardly to three or more pulses \cite{KoschorreckJPBAMOP2009}.
While a two-pulse experiment, plus prior knowledge of $\kappa$ and
$\tilde{J}_{3,3}$, gives sufficient information to find the
post-measurement variance, and thus test the state-preparation 
property, a three-pulse experiment is required to find
the other quantities used to characterize QND measurements.

If $\bST$ denotes the Stokes vector of the third probe pulse, then
statistics such as $\var(\ST_y)$ and $\cov(\SQ_y,\ST_y)$ can be
determined, and these in turn provide enough constraints to
determine the loss and noise.  Expanding our system to $\bSJQV
\equiv \bJ \oplus \bSQ \oplus \bSV \oplus \bST$, and defining
interaction and noise operators $M_\ST,N_\ST$ in the obvious way,
a direct calculation finds several useful relations \bea \AStay &=&
\frac{\delta\cov(\SQ_y,\ST_y)}{\delta\cov(\SQ_y,\SV_y)}  
\\  r_A^2 &=& { \frac{\delta\var(\ST_y) - \delta\var(\SV_y)}
{\delta\var(\SV_y) - \delta\var(\SQ_y)}} \\
\kappa^2 N_{3,3} &=& \delta\var(\SV_y) - \delta\var(\SQ_y) +
\kappa^2 \tilde{J}_{3,3}
\left(1-\AStay^2 \right)  \\
\kappa N_{3,5} &=& \delta\cov(\SQ_y,\SV_y) - \kappa^2 \tilde{J}_{3,3} \AStay\\
N_{5,5} &=& \delta\var(\SQ_y) - \kappa^2 \tilde{J}_{3,3}. \eea

\section{Measures of QND performance}


\newcommand{\HS}{{X}}
\newcommand{\HM}{{Y}}

To quantify QND performance, Holland {\em et al.} use the degree of correlation between various
combinations of the input and output system variable $\HS = J_z$ and meter variable $\HM = S_y$
variables  \cite{holland90}.  They define three figures of merit, each of which is unity for an ideal QND measurement.  These describe the
measurement quality, the preservation of the initial value, and the state 
preparation capability, respectively: 
\begin{eqnarray}
C^2_{\HS^{\rm in},\HM^{\rm out}}  &\equiv& \frac{\cov^2(\HS^{\rm in},\HM^{\rm out})}{\var(\HS^{\rm in})\var(\HM^{\rm out})}  =  \frac{\kappa^2
\tilde{J}_{3,3}^2}{\tilde{J}_{3,3} (\tilde{\SJQV}_\SQ)_{5,5}} =
\frac{\kappa^2 \tilde{J}_{3,3}}{\var(\SQy)} \\ C^2_{\HS^{\rm
in},\HS^{\rm out}} &\equiv& \frac{\cov^2(\HS^{\rm in},\HS^{\rm out})}{\var(\HS^{\rm in})\var(\HS^{\rm out})}  = \frac{\AStay^2
\tilde{J}_{3,3}^2}{\tilde{J}_{3,3} (\tilde{\SJQV}_\SQ)_{3,3}}  \nonumber \\ 
& = &
\frac{\kappa^2 \tilde{J}_{3,3} \delta \cov^2(\SQy,\STy)}{\delta
\cov^2(\SQy,\SVy) [\delta\var(\SVy)-\delta\var(\SQy) + \kappa^2
\tilde{J}_{3,3}]} \\ 
C^2_{\HS^{\rm out},\HM^{\rm out}} &\equiv& \frac{\cov^2(\HS^{\rm out},\HM^{\rm out})}{\var(\HS^{\rm out})\var(\HM^{\rm out})} =
\frac{(\tilde{\SJQV}_\SQ)_{3,5}^2}{(\tilde{\SJQV}_\SQ)_{3,3}
(\tilde{\SJQV}_\SQ)_{5,5}} \nonumber \\ & = & \frac{\delta
\cov^2(\SQy,\SVy)}{\var(\SQy)[\delta\var(\SVy)-\delta\var(\SQy) +
\kappa^2 \tilde{J}_{3,3}]}  \\ & &  \qtext{\rm\cov\rightarrow\cov^2~in~definitions.} \nonumber \end{eqnarray}




\section{Non-classicality criteria}

Roch, {\em et al.} \cite{RochAPBLO1992} and Grangier {\em et al.} \cite{grangier92}  define non-classicality criteria using the conditional
variance $\Delta X_{s|m}^2$, as in Eq. (\ref{Eq:EffectivenessWLoss}), and the quantities $\Delta X_m^2$, the measurement noise referred to the input and $\Delta X_s^2$, the excess noise introduced into the system variable. All are normalized by the intrinsic quantum noise of the system variable, a quantity which may depend on the system or the application.  For example, in a spin-squeezing context the natural noise scale is $\tilde{J}_{0} =  |\expect{J_x}| /2 = \tilde{J}_{3,3}$, the $J_z$ variance of  the input $x$-polarized coherent spin state, i.e., the projection noise.  Here we choose to normalize  $\Delta X_m^2$ by  $\tilde{J}_{0}$, and $\Delta X_{s|m}^2,\Delta X_s^2$ by $r_A \tilde{J}_{0}$, reflecting the reduction in size of the spin due to losses in the measurement process.  The relation of information gained to damage caused is non-classical if $\Delta X_s\Delta X_m < 1$. We find 
\bea
\Delta X_{s|m}^2 & \equiv & \frac{E[\var(J_z)|P_y]}{r_A \tilde{J}_0}  \qtext{\rm~fixed \times 2.~c.f.~Eqs~27~and~29}   \nonumber \\
&  =&  \frac{\delta
\cov(\SQy,\SVy)}{\delta
\cov(\SQy,\STy)} \left[1 
+  (\kappa^{2}\tilde{J}_{0})^{-1} \left( \delta \var(\SVy) - \delta \var(\SQy)-\frac{\delta\cov^2(\SQy,\SVy)}{ \var(\SQy)}   \right) \right]
\\
\Delta X_m^2 &\equiv& \frac{\tilde{C}_{2,2} \LTrans^2 +
N_{5,5}}{\kappa^2 \tilde{J}_{0} }
= \frac{\var(\SQy)-\kappa^2 \tilde{J}_{3,3}}{\kappa^2
\tilde{J}_{0}}  \\ \Delta X_s^2 &\equiv& \frac{
(\tilde{\SJQV}_\SQ)_{3,3}- \tilde{J}_{3,3}}{\AStay \tilde{J}_{0}
} = \frac{\delta
\cov(\SQy,\SVy)[\delta\var(\SVy)-\delta\var(\SQy)]}{\delta
\cov(\SQy,\STy)\kappa^2 \tilde{J}_{0}}.  \eea

  
\section{Conclusions}
Using the covariance matrix formalism and a general noise model, we have shown that full certification of
QND measurements is possible without direct access to the system variable under study.  We find that repeated probing of the same system gives statistical information sufficient to quantify both
 the state preparation capability and the information-damage tradeoff.  The results enable 
certification of true quantum non-demolition measurement of material systems, and are 
 directly applicable to ongoing experiments using QND measurements for quantum information \cite{echaniz08} and quantum-enhanced metrology \cite{windpassingerPRL2008,KoschorreckPRL2010a,KoschorreckPRL2010b}.

\section{Additional material}

The calculations described in this article can be performed in {\em Mathematica} using the notebook ``ThreePulseCMCalculator,'' available as an ancillary file.

\section{Acknowledgements}
We thank G. T\'{o}th for helpful discussions.  This work was supported by the 
Spanish MINECO under the project MAGO (Ref. FIS2011-23520) and by the European Research Council  under the project AQUMET.
\\~\\

\bibliographystyle{h-physrev}
\bibliography{QNDEffectiveness}

\begin{thebibliography}{10}

\bibitem{braginsky75}
V.~B. Braginsky and Y.~I. Vorontsov,
\newblock Sov. Phys. Usp. {\bf 17}, 644 (1975).

\bibitem{ThornePRL1978}
K.~S. Thorne, R.~W.~P. Drever, C.~M. Caves, M.~Zimmermann, and V.~D. Sandberg,
\newblock Phys. Rev. Lett. {\bf 40}, 667 (1978).

\bibitem{UnruhPRD1979}
W.~G. Unruh,
\newblock Phys. Rev. D {\bf 19}, 2888 (1979).

\bibitem{CavesRMP1980}
C.~M. Caves, K.~S. Thorne, R.~W.~P. Drever, V.~D. Sandberg, and M.~Zimmermann,
\newblock Rev. Mod. Phys. {\bf 52}, 341 (1980).

\bibitem{BraginskyS1980}
V.~B. Braginsky, Y.~I. Vorontsov, and K.~S. Thorne,
\newblock Science {\bf 209}, 547 (1980).

\bibitem{GrangierN1998}
P.~Grangier, J.~A. Levenson, and J.~P. Poizat,
\newblock Nature {\bf 396}, 537 (1998).

\bibitem{kuzmich99}
A.~Kuzmich {\em et~al.},
\newblock Phys. Rev. A {\bf 60}, 2346 (1999).

\bibitem{Takano2009}
T.~Takano, M.~Fuyama, R.~Namiki, and Y.~Takahashi,
\newblock Phys. Rev. Lett. {\bf 102}, 033601 (2009).

\bibitem{Appel2009}
J.~Appel {\em et~al.},
\newblock Proc. Natl. Acad. Sci. U.S.A. {\bf 106}, 10960 (2009).

\bibitem{Schleier-Smith2010a}
M.~H. Schleier-Smith, I.~D. Leroux, and V.~Vuleti\ifmmode~\acute{c}\else
  \'{c}\fi{},
\newblock Phys. Rev. Lett. {\bf 104}, 073604 (2010).

\bibitem{Chen2011}
Z.~Chen, J.~G. Bohnet, S.~R. Sankar, J.~Dai, and J.~K. Thompson,
\newblock Phys. Rev. Lett. {\bf 106}, 133601 (2011).

\bibitem{RuskovPR2005}
R.~Ruskov, K.~Schwab, and A.~Korotkov,
\newblock Phys. Rev. B {\bf 71}, 235407 (2005).

\bibitem{holland90}
M.~J. Holland, M.~J. Collett, D.~F. Walls, and M.~D. Levenson,
\newblock Phys. Rev. A {\bf 42}, 2995 (1990).

\bibitem{RochAPBLO1992}
J.~F. Roch, G.~Roger, P.~Grangier, J.-M. Courty, and S.~Reynaud,
\newblock Applied Physics B: Lasers and Optics {\bf 55}, 291 (1992).

\bibitem{grangier92}
P.~Grangier, J.~M. Courty, and S.~Reynaud,
\newblock Opt. Commun. {\bf 89}, 99 (1992).

\bibitem{BencheikhPRL1995}
K.~Bencheikh, J.~A. Levenson, P.~Grangier, and O.~Lopez,
\newblock Phys. Rev. Lett. {\bf 75}, 3422 (1995).

\bibitem{BencheikhPRL1997}
K.~Bencheikh, C.~Simonneau, and J.~A. Levenson,
\newblock Phys. Rev. Lett. {\bf 78}, 34 (1997).

\bibitem{BruckmeierPRL1997}
R.~Bruckmeier, H.~Hansen, and S.~Schiller,
\newblock Phys. Rev. Lett. {\bf 79}, 1463 (1997).

\bibitem{GoobarPRL1993}
E.~Goobar, A.~Karlsson, and G.~Bj\"ork,
\newblock Phys. Rev. Lett. {\bf 71}, 2002 (1993).

\bibitem{LevensonPRL1993}
J.~A. Levenson {\em et~al.},
\newblock Phys. Rev. Lett. {\bf 70}, 267 (1993).

\bibitem{PereiraPRL1994}
S.~F. Pereira, Z.~Y. Ou, and H.~J. Kimble,
\newblock Phys. Rev. Lett. {\bf 72}, 214 (1994).

\bibitem{PoizatPRL1993}
J.~P. Poizat and P.~Grangier,
\newblock Phys. Rev. Lett. {\bf 70}, 271 (1993).

\bibitem{RochPRL1993}
J.-F. Roch, J.-P. Poizat, and P.~Grangier,
\newblock Phys. Rev. Lett. {\bf 71}, 2006 (1993).

\bibitem{RochPRL1997}
J.-F. Roch {\em et~al.},
\newblock Phys. Rev. Lett. {\bf 78}, 634 (1997).

\bibitem{SewellARX2011}
R.~J. Sewell {\em et~al.},
\newblock (2011), 1111.6969v2.

\bibitem{kuzmich98}
A.~Kuzmich, N.~P. Bigelow, and L.~Mandel,
\newblock Europhys. Lett. {\bf 42}, 481 (1998).

\bibitem{Madsen:2004}
L.~B. Madsen and K.~M{\o}lmer,
\newblock Phys. Rev. A {\bf 70} (2004).

\bibitem{echaniz08}
S.~R. de~Echaniz, M.~Koschorreck, M.~Napolitano, M.~Kubasik, and M.~W.
  Mitchell,
\newblock Phys. Rev. A {\bf 77}, 032316 (2008).

\bibitem{KubasikPRA2009}
M.~Kubasik {\em et~al.},
\newblock Phys. Rev. A {\bf 79}, 043815 (2009).

\bibitem{KoschorreckPRL2010a}
M.~Koschorreck, M.~Napolitano, B.~Dubost, and M.~W. Mitchell,
\newblock Phys. Rev. Lett. {\bf 104}, 093602 (2010).

\bibitem{KoschorreckJPBAMOP2009}
M.~Koschorreck and M.~W. Mitchell,
\newblock Journal of Physics B: Atomic, Molecular and Optical Physics {\bf 42},
  195502 (2009).

\bibitem{windpassingerPRL2008}
P.~J. Windpassinger {\em et~al.},
\newblock Phys. Rev. Lett. {\bf 100}, 103601 (2008).

\bibitem{KoschorreckPRL2010b}
M.~Koschorreck, M.~Napolitano, B.~Dubost, and M.~W. Mitchell,
\newblock Phys. Rev. Lett. {\bf 105}, 093602 (2010).

\end{thebibliography}



\end{document}